\documentclass[article,twocolumn]{revtex4}
\usepackage{graphicx}
\usepackage{amssymb}
\usepackage{bm}
\usepackage{hyperref}
\usepackage{epstopdf}
\usepackage{color}
\usepackage{mathrsfs}

\begin{document}

\title{Exploring the low redshift universe: two parametric models for effective pressure}

\author{Qiang Zhang$^{1}$ Guang Yang$^{1}$}
\email{yang-guang@mail.nankai.edu.cn}
\author{Qixiang Zou$^{1}$ Xinhe Meng$^{1,2}$}
\email{xhm@nankai.edu.cn}
\author{Keji Shen$^{1}$}

\affiliation{
$^1$Department of Physics, Nankai University, Tianjin 300071, China\\
$^2$State Key Laboratory of Theoretical Physics China,CAS, Beijing 100190, China\\
{xhm@nankai.edu.cn(correspondence)}}

\date{Feb 5th 2015}

\begin{abstract}
Astrophysical observations have put unprecedentedly tight constraints on cosmological theories. The $\Lambda$CDM model, mathematically simple and fits observational data-sets well, is preferred for explaining the behavior of universe. But many basic features of the dark sectors are still unknown, which leaves rooms for various nonstandard cosmological hypotheses. As the pressure of cosmological constant dark energy is unvarying, ignoring contributions from radiation and curvature terms at low redshift, the effective pressure keeps constant. In this paper, we propose two parametric models for non-constant effective pressure in order to study the tiny deviation from $\Lambda$CDM at low redshift. We recover our phenomenological models in the scenarios of quintessence and phantom fields, and explore the behavior of scalar field and potential. We constrain our model parameters with SNe Ia and BAO observations, and detect subtle hints of $\omega_{de}<-1$ from the data fitting results of both models, which indicates possibly a phantom dark energy scenario at present.
\end{abstract}

\maketitle
\section{Introduction}

Since the discovery of current acceleration of our universe expansion in 1998, maybe the greatest mystery in cosmology is the deceptive nature of the dark energy. Recent observational results~\cite{Planck} have put tight constraints on the properties of dark energy, but there is still no theoretical or observational indication to pin down its nature. On one hand, although the simple cosmological constant $\Lambda$ can accommodate the accelerating expansion, it encounters two serious problems. The first one is the fine tuning problem: the measured energy of the vacuum so much smaller than the estimated value $\rho_{vac}^{obs}\ll\rho_{vac}^{theo}$, which is the famous 120-orders-of-magnitude discrepancy that makes the vacuum explanation suspecious. The second one is why the dominance of the cosmological constant over the matter component at the present epoch. These two basic problems prompt us to propose some alternatives, which include an evolving scalar field called quintessence~\cite{quintessence1,quintessence2,quintessence3,quintessence4,quintessence5,modifiedgravity00,modifiedgravity000}, noncanonical scalar field (such as K-essence~\cite{K-essence1,K-essence2,K-essence3}, phantom~\cite{modifiedgravity00,modifiedgravity000,phantom1,phantom1',phantom1'',phantom1''',phantom3,phantom4,phantom5}), modified gravity~\cite{modifiedgravity00,modifiedgravity000,modifiedgravity0,modifiedgravity1,modifiedgravity2,
modifiedgravity3,modifiedgravity4}, coupled dark energy~\cite{modifiedgravity000,coupledDE1,coupledDE2} or decaying dark energy~\cite{decayingDE} and so on. On the other hand, as we know the equation of state (EoS) parameter of the cosmological constant is precisely $\omega_{de}=-1$. Recent observations show that the EoS parameter of modeled dark energy is $\omega_{de}=-1.006\pm 0.045$, which slightly favours $\omega_{de}<-1$. Anyhow, the small deviations from the cosmological constant $\Lambda$ allow one to consider models with $\omega_{de}\neq-1$. So one can make efforts to construct new models to explain the deviations which may be detectable at the precision of current and future observations.

Parameterization is an useful tool towards a more complete characterization of dark energy modelling and has been routinely employed to analyze datasets. Most parameterizations for dark energy models involve the EoS parameter $\omega_{de}$ for the dark energy behavior. Several well-known parameterizations for the EoS of dark energy have been proposed so far. We can write parameterizations in polynomial form $\omega_{de}(z)=\sum\limits_{n=0}\omega_nx_n(z)$ generally, where the expansions can be given by the following, (i)Redshift: $x_n(z)=z^n$, (ii)Scale factor: $x_n(z)=(1-\frac{a}{a_0})^n=(\frac{z}{1+z})^n$, (iii)Logarithmic: $x_n(z)=[\ln(1+z)]^n$. Parameterization (i) was proposed by Huterer and Turner\cite{i1} and Weller and Albrecht\cite{i2} with $n\leq1$. Parameterization (ii) with $n\leq1$ was introduced by Chevalier,  Polarski and Linder~\cite{CPLa,CPLb}, the famous Chevallier-Polarski-Linder(CPL) parameterization. $\omega_{de}=\omega_0+\omega_1(1-a)=\omega_0+\omega_1\frac{z}{1+z}$ behaves as $\omega_{de}\rightarrow\omega_0+\omega_1$ for $z\rightarrow\infty$ and $\omega_{de}\rightarrow\omega_0$ for $z\rightarrow0$. A more general form with $\omega_{de}=\omega_0+\omega_1\frac{z}{(1+z)^p}$ was later proposed by Jassal, Bagla and Padmanabhan\cite{i3'}. Parameterization (iii) with $n\leq1$ was introduced by Efstathiou\cite{i3}. In recent years, some new parameterizations have been proposed, such as using Pad\'{e} parameterizations for the EoS of dark energy\cite{i4}, namely $\omega_{de}=\frac{\omega_0+\omega_a(1-a)}{1+\omega_b(1-a)}$ ,and $\omega_{de}=\frac{\omega_0+\omega_1\ln a}{1+\omega_2\ln a}$. It is worth mentioning that Sen proposed a parameterization for the pressure of dark energy model~\cite{Sen,Sumit}, $P_\Lambda=-P_0+P_1(1-a)+\cdot\!\cdot\!\cdot\cdot $, in order to study small deviations from the cosmological constant. Different from parameterizations which focused on the EoS of dark energy mentioned above, in this paper we aim to make parameterizations for the relation between redshift and effective pressure of all energy components in the universe. In the following we proposed two parametric models for the effective pressure in order to explore late-stage evolution of the universe.

This paper is organized as follows: in section II, we propose two new parametric models for the effective pressure: $P(z)=P_a+P_b z$ and $P(z)=P_c+\frac{P_d}{1+z}$. In Section III, we relate our parametric models with the quintessence and phantom scalar fields, and the behavior of field and potential is then explored. In Section IV, we constrain our model parameters with SNe Ia and BAO observations. In Section V, We end with discussions and conclusions.

\section{Two parametric models }
The Friedmann equations, equation of energy conservation and equation of state constitute a close system to describe the background evolution of the universe. A substitute from EoS to a relation between effective pressure $P$ and redshift $z$ is also feasible, as equation $P=P(z)$ is not linearly dependent on the Friedmann equation and equation of energy conservation. Also, the EoS can be recovered by inserting $P-z$ relation into equation of energy conservation
\begin{equation}
 \dot{\rm \rho}+3 H(P+\rho)=0\label{eq1-1},
 \end{equation}
and integrating out the expression of $\rho$. For example, the effective pressure for $\Lambda CDM$ at late stage is nearly constant, say $P_0 $; accordingly, we can obtain from Eq.~(\ref{eq1-1}) that
 \begin{equation}
 \rho(a)=-P_0+C a^{-3} \label{eq1-11111}
 \end{equation}
where $C$ is an integration constant, and the two terms at the right side represent contributions from cosmological constant and matter respectively.

This is just an example for $P$ parameterization; generally, we can have more complicated $P-z$ relations. As $P-z$ relation is equivalent to EoS, a parameterization on the effective pressure is equivalent to that on the EoS parameter $\omega_{de}$. Since $\omega_{de}$ is the exponential of some component in EoS, $\omega_{de}$ paramerterization requests a presupposition of the components in EoS; i.e., the physical mechanism of the possible deviation from $\Lambda$CDM has to be dictated although We make parameterizations merely because we actually do not know the concrete mechanism behind the accelerative expansion. To illustrate, a deviation of $\Lambda$CDM might come from the evolution of the E0S of the cosmological constant term, while an additional component can also result in same deviation. However, a parameterization on the effective pressure just circumvents this issue, and do not require any knowledge of the concrete physical mechanism. We are able to directly study the deviation from constant $P-z$ relation without prejudice to a presupposition.

\subsection{Model 1}
In this subsection, we propose a model which reads:

\begin{equation}
  P(z)=P_a+P_b z \label{eq1-2},
\end{equation}
where $P_a$ and $P_b$ are free parameters.

For scale factor $a$ and redshift $z$, we have
\begin{equation}
  a=\frac{a_0}{1+z}=\frac{1}{1+z}\label{eq1-3},
\end{equation}
where $a_0=1$ corresponds to the value today. Substitute Eqs.~(\ref{eq1-2})and ~(\ref{eq1-3}) to Eq.~(\ref{eq1-1}), the total energy density can be integrated as
 \begin{equation}
 \rho(a)=-(P_a-P_b)-\frac{3}{2}P_b a^{-1}+C_1 a^{-3} \label{eq1-4},
 \end{equation}
where $C_1$ is an integration constant. If we set $\rho_0$to be the energy density today,  the integration constant is then $C_1=\rho_0+P_a+\frac{1}{2}P_b$. In Eq.~(\ref{eq1-4}), we can interpret the inversely cubic term $C_1 a^{-3}$ as dust matter and the constant term $-(P_a-P_b)$ as the cosmological constant in $\Lambda$CDM. Term -$\frac{3}{2}P_b a^{-1}$ does not appear in the $\Lambda$CDM model, whose physical nature will be explored in next section.

For convenience in date fitting, we introduce some dimensionless parameters. First, we define dimensionless density and pressure as
\begin{eqnarray}
  \rho^\ast&\equiv&\frac{\rho}{\rho_0}=\frac{H^2}{H_0^2} \label{eq1-5}, \\
  P^\ast&\equiv&\frac{P}{\rho_0} \label{eq1-6}.
\end{eqnarray}

The expressions of the total density Eq.~(\ref{eq1-4}) and total pressure Eq.~(\ref{eq1-2}) can be rewritten as:
\begin{eqnarray}
  \rho^\ast(a)&=&-(P_a^\ast-P_b^\ast)-\frac{3}{2}P_b^\ast a^{-1}+C_1^\ast a^{-3} \label{eq1-7} ,\\
  P^\ast(a)&=&(P_a^\ast-P_b^\ast)+P_b^\ast a^{-1} \label{eq1-8},
\end{eqnarray}
where $P_a^\ast\equiv \frac{P_a}{\rho_0}$, $P_b^\ast\equiv \frac{P_b}{\rho_0}$ and $C^\ast_1\equiv \frac{C_1}{\rho_0}=1+P_a^\ast+\frac{1}{2}P_b^\ast$.

Redefined two new parameters $\alpha\equiv -(P_a^\ast-P_b^\ast)$ and $\beta\equiv -\frac{3}{2}P_b^\ast$, then:
\begin{eqnarray}
  \rho^\ast(a)&=&\alpha+\beta a^{-1}+(1-\alpha-\beta) a^{-3} \label{eq1-9} ,\\
  P^\ast(a)&=&-\alpha-\frac{2}{3}\beta a^{-1} \label{eq1-10}.
\end{eqnarray}

As we know, the dimensionless Hubble parameter is
\begin{equation}
    E(z)\equiv \frac{H}{H_0} \label{eq1} .
\end{equation}
Compare Eq.~(\ref{eq1}) with Eq.~(\ref{eq1-5}), we obtain,
\begin{equation}
    E(a)= \rho^\ast(a)^\frac{1}{2} \label{eq2}.
\end{equation}
Then, for model 1, we define
\begin{eqnarray}
  \Omega_1&=&\frac{\alpha}{E^2} \label{eq3}, \\
  \Omega_2&=&\frac{\beta a^{-1}}{E^2} \label{eq4},\\
  \Omega_m&=&\frac{\Omega_{m0}a^{-3}}{E^2} \label{eq5},
\end{eqnarray}
where $\Omega_{m0}=1-\alpha-\beta$, hence $\Omega_1+\Omega_2+\Omega_m=1$.

\subsection{Model 2}
We propose another parameterization as
\begin{equation}
 P(z)=P_c+\frac{P_d}{1+z} \label{eq1-11},
\end{equation}
where $P_c$ and $P_d$ are free parameters. Inserting Eqs.~(\ref{eq1-3}) and ~(\ref{eq1-11}) into Eq.~(\ref{eq1-1}), we obtain the total energy density for model 2,
\begin{equation}
\rho(a)=-P_c-\frac{3}{4}P_d a+C_2 a^{-3} \label{eq1-12},
\end{equation}
where $C_2$ is an integration constant. Set the present energy density as $\rho_0$, then $C_2=\rho_0+P_c+\frac{3}{4}P_d$. Still, we can find term $C_2 a^{-3}$ corresponding to dust matter, and term $-P_c$corresponding to the cosmological constant. The difference between model 2 and model 1 rests on the rest term, $-\frac{3}{2}P_b a^{-1}$ for model 2 whereas $-\frac{3}{4}P_d a$ for model 1.Their physical nature will be studied in next section.

Like model 1, we need to introduce new model parameters in model 2. With Eqs.~(\ref{eq1-5}) and ~(\ref{eq1-6}), we can obtain the expressions of total density and total pressure for model 2:

\begin{eqnarray}
  \rho^\ast(a)&=&-P_c^\ast-\frac{3}{4}P_d^\ast a+C_2^\ast a^{-3} \label{eq1-13}, \\
  P^\ast(a)&=&P_c^\ast+P_d^\ast a \label{eq1-14},
\end{eqnarray}

\noindent
where $P_c^\ast\equiv \frac{P_c}{\rho_0}$, $P_d^\ast\equiv \frac{P_d}{\rho_0}$ and $C^\ast_2\equiv \frac{C_2}{\rho_0}=1+P_c^\ast+\frac{3}{4}P_d^\ast$.

Redefine two new parameters $\gamma\equiv -P_c^\ast$ and $\delta\equiv -\frac{3}{4}P_d^\ast$, then:
\begin{eqnarray}
  \rho^\ast(a)&=&\gamma+\delta a+(1-\gamma-\delta) a^{-3} \label{eq1-15} ,\\
  P^\ast(a)&=&-\gamma-\frac{4}{3}\delta a \label{eq1-16}.
\end{eqnarray}

Also, we define for model 2,
\begin{eqnarray}
  \Omega_1&=&\frac{\gamma}{E^2} \label{eq6}, \\
  \Omega_2&=&\frac{\delta a}{E^2} \label{eq7} ,\\
  \Omega_m&=&\frac{\Omega_{m0}a^{-3}}{E^2} ,\label{eq8}
\end{eqnarray}
where $\Omega_{m0}=1-\gamma-\delta$, we have $\Omega_1+\Omega_2+\Omega_m=1$.

\section{Relation with scalar fields}
Deviations from the $\Lambda$CDM in our models can be realized through different physical scenarios. Scalar fields are mainstream approaches to explain the acceleration of the universe expansion. In the scenarios of scalar fields,  dark energy evolves with time. The scalar field dynamics has been studied by literature in great details(see Ref.~\cite{quintessence1,quintessence2,quintessence3,quintessence4,quintessence5,modifiedgravity00,modifiedgravity000,K-essence1,K-essence2,K-essence3,phantom1,phantom1',phantom1'',phantom1''',phantom3,phantom4,phantom5}) and there are lots of issues involved such as (i)choosing initial conditions for scalar field; (ii)choosing potential with solid theoretical motivation; (iii) the existence of the ¡°tracker¡± field and so on. Generally, evolution of scalar field is studied over the cosmic history, and once the parameters of scalar field models are set they determine the entire cosmological evolution. So a more detailed analysis would involve studying scalar field dynamics over cosmic history, and then comparing its evolution with that of pressure parametrization model at low redshift. In this paper, we will merely compare pressure and energy density of field with that of a model of pressure parametrization at low redshift and study the behavior of field and potential. Physical realization of parameterizations through scalar fields means adjusting behavior of scalar fields to dark energy term occurred in parametric model. Specifically speaking, we make two equations
\begin{eqnarray}
P_{eff} = P_{scalar field},\\
\rho_{eff}-\rho_m = \rho_{scalar field},
\end{eqnarray}
as mathematical definition of realization.

In this section, we will take ``quintessence'' and ``phantom'' as two examples to realize our models.

{\bf Quintessence}: ``Quintessence'' denotes a canonical scalar field $\phi$ with a potential $V_1(\phi)$ that does nor interact with all the other components except standard gravity, whose EoS parameter $\omega_{de}>-1$. The quintessence is described by action
\begin{eqnarray}
  S &=& \int d^4x\sqrt{-g}[\frac{1}{2\kappa^2}R+\mathscr{L}_\phi]+S_M,    \\
  \mathscr{L}_\phi &=& -\frac{1}{2}g^{\mu\nu}\partial_{\mu}\phi\partial_{\nu}\phi-V_1(\phi) \label{eq3-11},
\end{eqnarray}
where $\kappa^2=8\pi G$, $R$ is the Ricci scalar and $S_M$ is the action of matter. The variation of the action Eq.~(\ref{eq3-11}) with respect to $\phi$ gives
\begin{equation}
\ddot{\phi} + 3 H \dot{\phi} + V^\prime_1(\phi) = 0\label{eq3-12},
\end{equation}
where $V_1(\phi)$ is the potential of the quintessence field, the prime denotes the derivative with respect to $\phi$. In a FLRW background, the energy density $\rho_{de}$ and the pressure $P_{de}$ of the quintessence field are
\begin{eqnarray}
 \rho_{de} &=& \frac{1}{2}\dot{\phi}^2+V_1(\phi)\label{eq3-13}, \\
 P_{de} &=& \frac{1}{2}\dot{\phi}^2-V_1(\phi)\label{eq3-14}.
\end{eqnarray}
Then the EoS
\begin{equation}
 \omega_{de}=\frac{\frac{1}{2}\dot{\phi}^2-V_1(\phi)}{\frac{1}{2}\dot{\phi}^2+V_1(\phi)}\label{eq3-15}.
\end{equation}

{\bf Phantom}: Minimally coupled phantom model is also a possible realization, whose EoS parameter $\omega_{de}<-1$. The action of the phantom field minimally coupled to gravity and matter sources is
\begin{eqnarray}
  S &=& \int d^4x\sqrt{-g}[\frac{1}{2\kappa^2}R+\mathscr{L}_\phi]+S_M,    \\
  \mathscr{L}_\phi &=& \frac{1}{2}g^{\mu\nu}\partial_{\mu}\phi\partial_{\nu}\phi-V_2(\phi) ,\label{eq3-16}
 \end{eqnarray}
whose variation with respect to $\phi $ gives
 \begin{equation}
  \ddot{\phi} + 3 H \dot{\phi} - V^\prime_2(\phi) = 0 ,\label{eq3-17}
 \end{equation}
 \noindent
where $V_2(\phi)$ is the potential of the phantom field, and the prime denotes the derivative with respect to $\phi$. The energy density and pressure of the phantom are given by(assuming flat FRW metric)
\begin{eqnarray}
 \rho_{de} &=&-\frac{1}{2}\dot{\phi}^2+V_2(\phi)  \label{eq3-18} ,\\
 P_{de} &=& -\frac{1}{2}\dot{\phi}^2-V_2(\phi) \label{eq3-19}.
\end{eqnarray}

The EoS of the phantom field is then
\begin{equation}
 \omega_{de}=-\frac{-\frac{1}{2}\dot{\phi}^2-V_2(\phi)}{-\frac{1}{2}\dot{\phi}^2+V_2(\phi)}\label{eq3-20}.
\end{equation}
So $\omega_{de}<-1$ for $\frac{1}{2}\dot{\phi}^2<V_2(\phi)$.

\subsection{Model 1}

The EoS of the scalar fields for model 1 reads
\begin{equation}
 \omega_{de}=\frac{ P_{scalar field}}{\rho_{scalar field}}=-1+\frac{ \frac{1}{3}\beta(1+z)}{\alpha+\beta(1+z)} \label{eq3-21}.
 \end{equation}
\noindent
Note that in above equation, there will be a singularity when $z=-\frac{\alpha}{\beta}-1$. In this paper we only consider the universe at low redshift, so we need not to worry about that situation. Besides in section IV data fitting will support our argument.

In the quintessence scenario, assuming the cosmic components consist of matter and quintessence, comparing Eq.~(\ref{eq3-13}) and Eq.~(\ref{eq3-14}) with Eq.~(\ref{eq1-4}) and Eq.~(\ref{eq1-2}), we have
\begin{eqnarray}
 -(P_a-P_b)-\frac{3}{2}P_b a^{-1} &=& \frac{1}{2}\dot{\phi}^2+V_1(\phi) \label{eq3-22} ,\\
 P_a-P_b+P_b a^{-1} &=& \frac{1}{2}\dot{\phi}^2-V_1(\phi) \label{eq3-23}.
\end{eqnarray}

Simplify the above two equations, compare to Eqs.~(\ref{eq1-5})~(\ref{eq1-6})~(\ref{eq1-7})~(\ref{eq1-8})~(\ref{eq1-9})~(\ref{eq1-10}), replace model parameters ($P_a$ , $P_b$) with redefined parameters ($\alpha$ , $\beta$), then we obtain
\begin{eqnarray}
\frac{1}{2}\dot{\phi}^2 &=& \frac{1}{6}\rho_0\beta a^{-1} \label{eq3-24} ,\\
 V_1(\phi) &=& \rho_0\alpha+\frac{5}{6}\rho_0\beta a^{-1} \label{eq3-25}.
\end{eqnarray}

From Eq.~(\ref{eq3-24}), it is easy to find that $\beta>0$ in the scenario of quintessence. By Eq.~(\ref{eq3-24}) and Eq.~(\ref{eq3-25}), one can construct the kinetic energy $\frac{1}{2}\dot{\phi}^2$ and potential $V_1(\phi)$ of the quintessence field with parameters ($\alpha$ , $\beta$) of model 1. In order to solve the above two equations, following~\cite{carroll}, we choose condition $\phi_{a=1}=M_P$, where $M_P$ is the reduced Planck mass. The Friedmann equation can then be rewritten as
\begin{equation}
 H^2=\frac{1}{3M_P^2}\rho \label{eq3-25'}.
\end{equation}

Considering the dark energy domination at present epoch in the universe, with the density parameter in dark energy $\Omega_{de}\sim0.7$, we define $V_0=\rho_0=3M_P^2H_0^2$. Simplify Eq.~(\ref{eq3-24}) and Eq.~(\ref{eq3-25}), we have
\begin{eqnarray}
\frac{d\phi}{da} &=& \pm M_P\sqrt{\frac{\beta}{\alpha+\beta a^{-1}+(1-\alpha-\beta) a^{-3}}} a^{-\frac{3}{2}} \label{eq3-25''}, \\
V_1(\phi) &=& V_0(\alpha+\frac{5}{6}\beta a^{-1}) \label{eq3-25'''}.
\end{eqnarray}

\begin{figure}[h!]
\begin{center}
\includegraphics[width=0.45\textwidth]{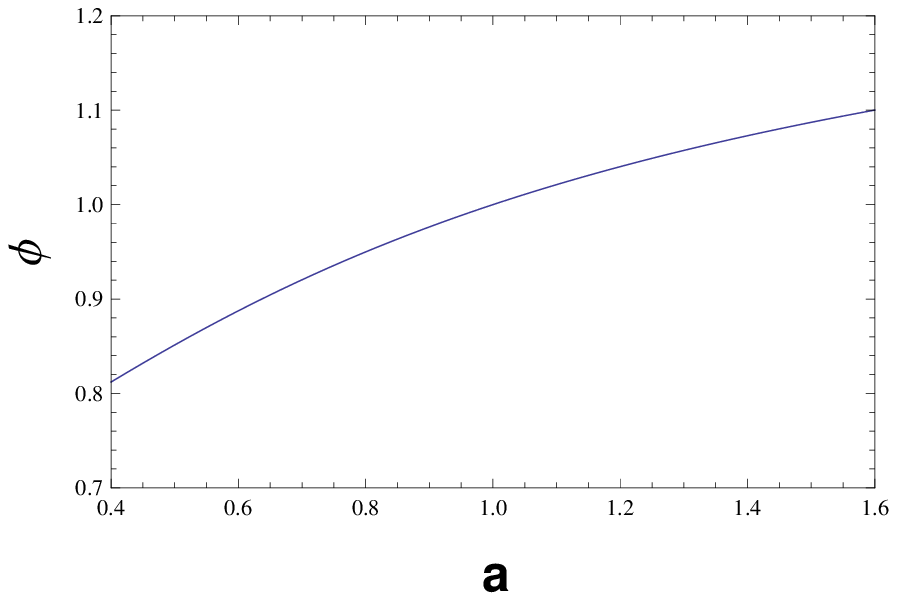}
\includegraphics[width=0.45\textwidth]{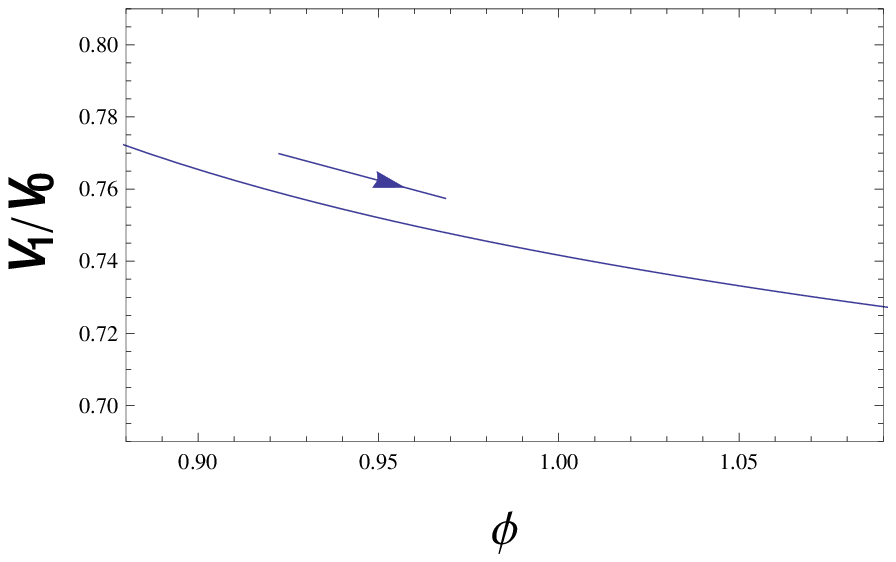}
\end{center}
\caption{The solution of Eq.~(\ref{eq3-25''}) and Eq.~(\ref{eq3-25'''}) corresponding to a plus sign in Eq.~(\ref{eq3-25''}). Field $\phi$ as a function of $a$ depicted in the top panel, potential $V_1$ as a function of $\phi$ depicted in the bottom panel. The arrow indicates evolutional direction of potential with respect to time. We consider values $\alpha=0.7,\beta=0.05$.}

\label{fig:aaa}
\end{figure}

\begin{figure}
\begin{center}
\includegraphics[width=0.45\textwidth]{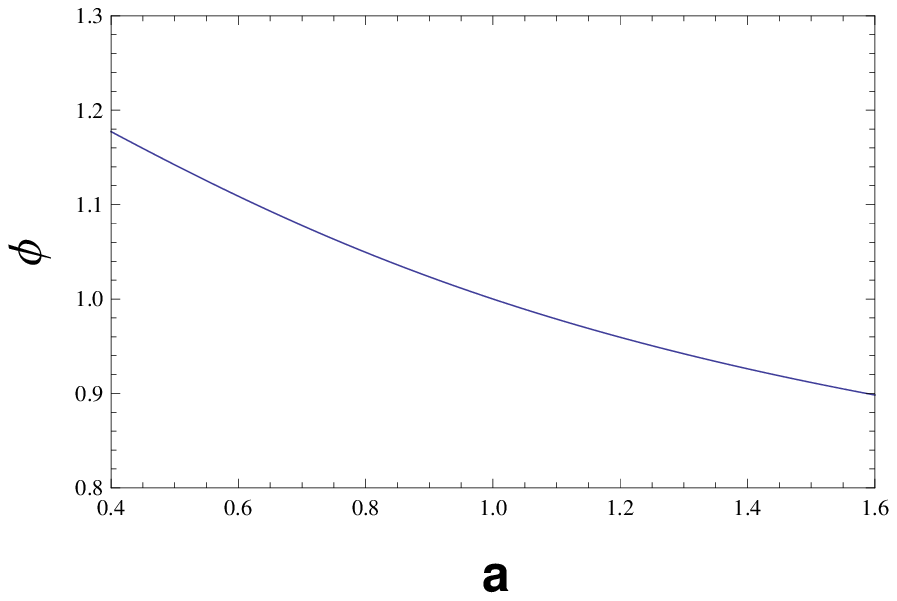}
\includegraphics[width=0.45\textwidth]{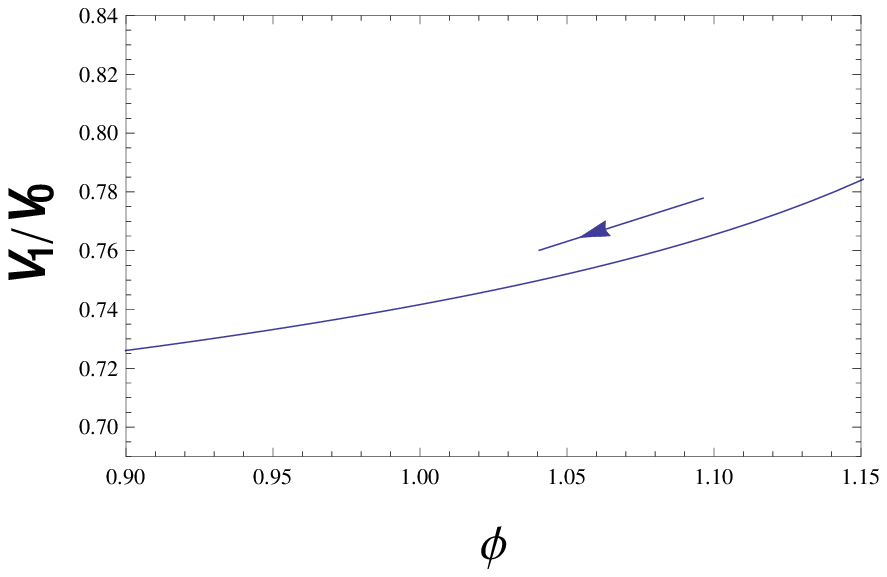}
\end{center}
\caption{The solution of Eq.~(\ref{eq3-25''}) and Eq.~(\ref{eq3-25'''}) corresponding to a minus sign in Eq.~(\ref{eq3-25''}). Field $\phi$ as a function of $a$ depicted in the top panel, potential $V_1$ as a function of $\phi$ depicted in the bottom panel. The arrow indicates evolutional direction of potential with respect to time. We consider values $\alpha=0.7,\beta=0.05$.}

\label{fig:bbbb}
\end{figure}

The symbol ``$\pm$'' in Eq.~(\ref{eq3-25''}) corresponds to two solutions. Consider $\alpha=0.7, \beta=0.05$ for numerically solving the above two equations, the solutions are represented in Fig.~(\ref{fig:aaa}) and Fig.~(\ref{fig:bbbb}) respectively. From Fig.~(\ref{fig:aaa}), we can find $\phi$ increases with $a$, the potential decreases with the increasing $\phi$, Eq.~(\ref{eq3-25'''}), implying that the potential will reach the minimum value $V_1(\phi)=V_0\alpha$ in the future. From Fig.~(\ref{fig:bbbb}), we can see that $\phi$ decreases with $a$, and the potential decreases with the decreasing $\phi$, Eq.~(\ref{eq3-25'''}) implies that the potential will reach the minimum $V_1(\phi)=V_0\alpha$ in the future.
By Eq.~(\ref{eq3}) and Eq.~(\ref{eq4}), we can obtain the expression of density parameter $\Omega_\phi$ for model 1:

\begin{equation}
 \Omega_\phi=\Omega_1+\Omega_2=\frac{\alpha+\beta a^{-1}}{E^2} \label{eq3-25-1}.
\end{equation}

\begin{figure}[h!]
\begin{center}
\includegraphics[width=0.40\textwidth]{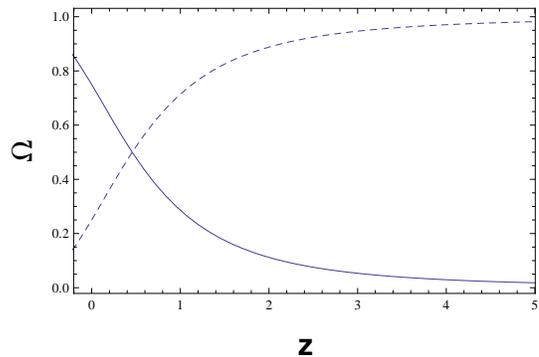}
\end{center}
\caption{Evolution of the density parameters in the quintessence field ($\Omega_\phi$) and matter ($\Omega_m$) for model 1. $\Omega_\phi$ is indicated by solid line, and $\Omega_m$ is indicated by dashed line. We consider values $\alpha=0.7,\beta=0.05$.}
\label{fig:1-quin}
\end{figure}

\noindent
The evolution of density parameter $\Omega_\phi$ in the scenario of quintessence is plotted in Fig.~(\ref{fig:1-quin}). From Fig.~(\ref{fig:1-quin}), we can see that until low redshift the energy density in the quintessence field becomes cosmologically dominant. Finally, the field comes to rest at the minimum of the potential $V_1(\phi)=V_0\alpha$, and the universe eventually settles into a de Sitter phase (see Eq.~(\ref{eq3-21})).

In the case of phantom scenario, assuming the cosmological components consist of matter and phantom, comparing Eq.~(\ref{eq3-18}) and Eq.~(\ref{eq3-19}) with Eq.~(\ref{eq1-4}) and Eq.~(\ref{eq1-2}), then we have

\begin{eqnarray}
 -(P_a-P_b)-\frac{3}{2}P_b a^{-1} &=& -\frac{1}{2}\dot{\phi}^2+V_2(\phi) \label{eq3-26} ,\\
 P_a-P_b+P_b a^{-1} &=& -\frac{1}{2}\dot{\phi}^2-V_2(\phi) \label{eq3-27}.
\end{eqnarray}

Replace model parameters ($P_a$ , $P_b$) with redefined parameters ($\alpha$ , $\beta$), we have

\begin{eqnarray}
\frac{1}{2}\dot{\phi}^2 &=& -\frac{1}{6}\rho_0\beta a^{-1} \label{eq3-28} ,\\
 V_2(\phi) &=& \rho_0\alpha+\frac{5}{6}\rho_0\beta a^{-1} \label{eq3-29}.
\end{eqnarray}
From Eq.~(\ref{eq3-28}), it is easy to find that in the scenario of phantom, $\beta<0$. By Eq.~(\ref{eq3-28}) and Eq.~(\ref{eq3-29}), one can construct the kinetic energy $\frac{1}{2}\dot{\phi}^2$ and potential $V_2(\phi)$ of the phantom field with model parameter ($\alpha$ , $\beta$). Eq.~(\ref{eq3-28}) and Eq.~(\ref{eq3-29}) can be rewritten as

\begin{eqnarray}
\frac{d\phi}{da} &=& \pm M_P\sqrt{\frac{-\beta}{\alpha+\beta a^{-1}+(1-\alpha-\beta) a^{-3}}} a^{-\frac{3}{2}} \label{eq3-29'}, \\
 V_2(\phi) &=& V_0(\alpha+\frac{5}{6}\beta a^{-1}) \label{eq3-29''}.
\end{eqnarray}

\begin{figure}
\begin{center}
\includegraphics[width=2.6in,height=2in,angle=0]{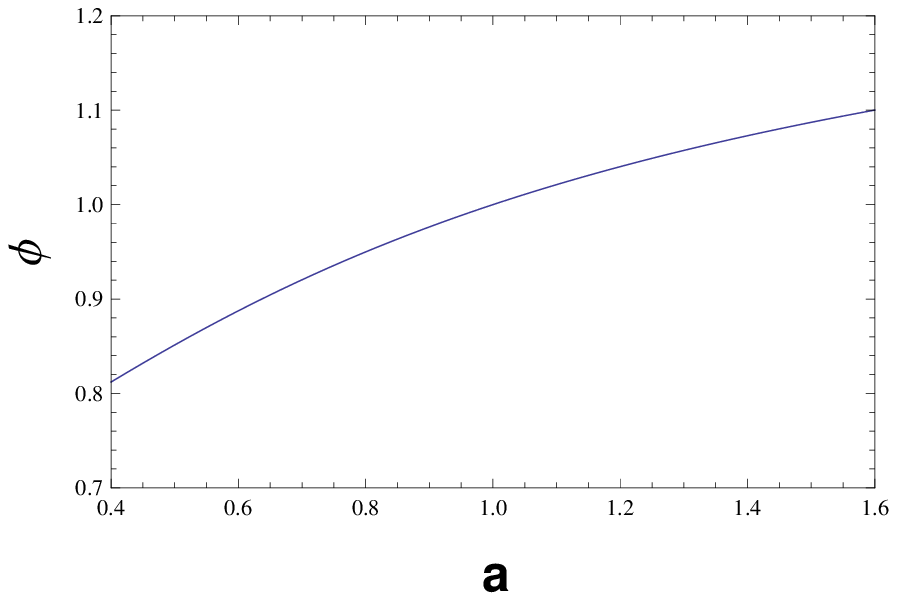}
\includegraphics[width=2.6in,height=2in,angle=0]{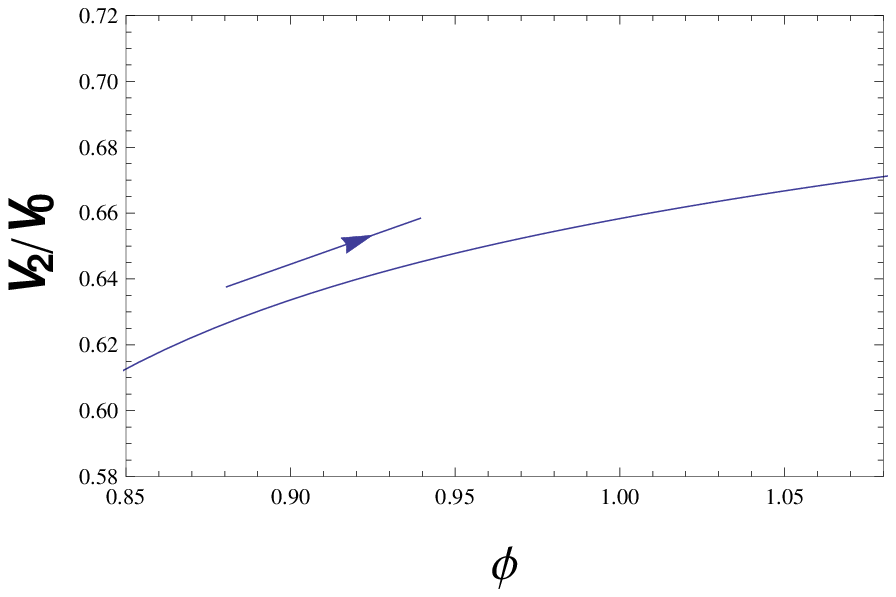}
\end{center}
\caption{The solution of Eq.~(\ref{eq3-29'}) and Eq.~(\ref{eq3-29''}) corresponding to a plus sign in Eq.~(\ref{eq3-29'}). Field $\phi$ as a function of $a$ depicted in the top panel, potential $V_2$ as a function of $\phi$ depicted in the bottom panel. The arrow indicates evolutional direction of potential with respect to time. We consider values $\alpha=0.7,\beta=-0.05$.}

\label{fig:ccc}
\end{figure}

\begin{figure}
\begin{center}
\includegraphics[width=2.6in,height=2in,angle=0]{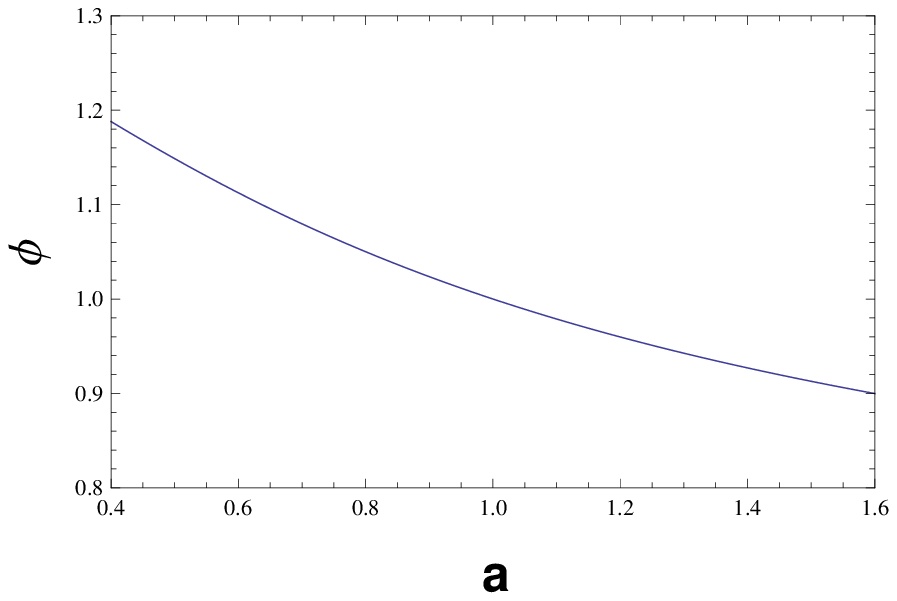}
\includegraphics[width=2.6in,height=2in,angle=0]{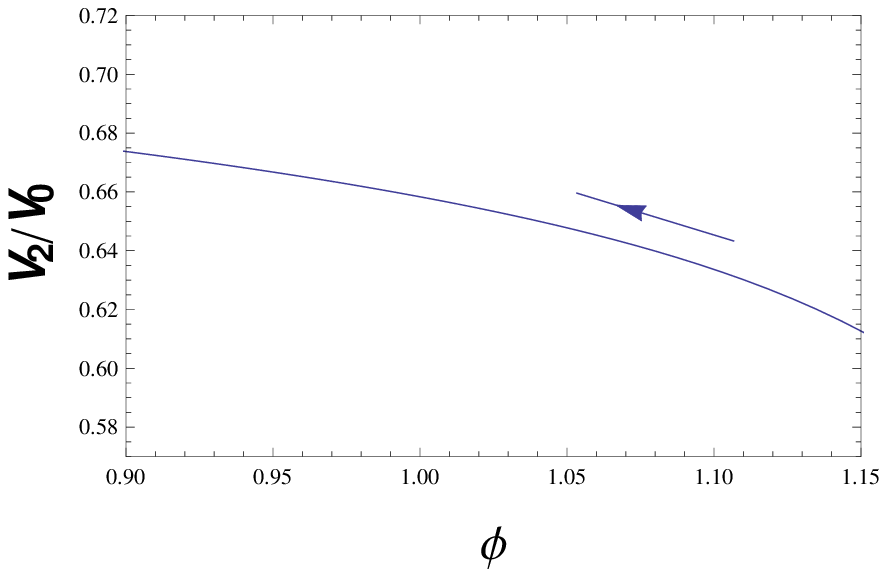}
\end{center}
\caption{The solution of Eq.~(\ref{eq3-29'}) and Eq.~(\ref{eq3-29''}) corresponding to a minus sign in Eq.~(\ref{eq3-29'}). Field $\phi$ as a function of $a$ depicted in the top panel, potential $V_2$ as a function of $\phi$ depicted in the bottom panel. The arrow indicates evolutional direction of potential with respect to time. We consider values $\alpha=0.7,\beta=-0.05$.}

\label{fig:ddd}
\end{figure}

\begin{figure}[h!]
\begin{center}
\includegraphics[width=0.40\textwidth]{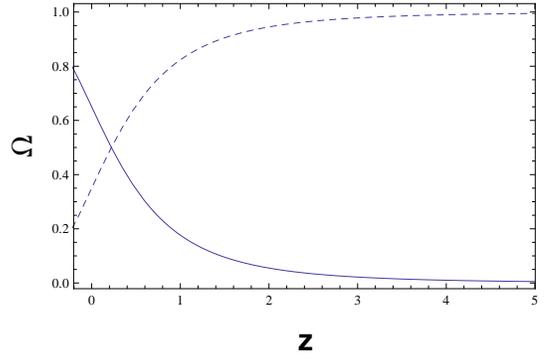}
\end{center}
\caption{Evolution of the density parameters in the phantom field ($\Omega_\phi$) and matter ($\Omega_m$) for model 1. $\Omega_\phi$ is indicated by solid line, and $\Omega_m$ is indicated by dashed line. We consider values $\alpha=0.7,\beta=-0.05$.}
\label{fig:1-phantom}
\end{figure}

Consider $\alpha=0.7, \beta=-0.05$ for numerically solving the above two equations, the two solutions are represented in Fig.~(\ref{fig:ccc}) and Fig.~(\ref{fig:ddd}) respectively. From Fig.~(\ref{fig:ccc}), we can find $\phi$ increases with $a$, and the potential increases with the increasing $\phi$, Eq.~(\ref{eq3-29''}) implies that the potential will reach the maximum value $V_2(\phi)=V_0\alpha$ in the future. From Fig.~(\ref{fig:ddd}), $\phi$ decreases with $a$, and the potential increases with the decreasing $\phi$, in the future the potential will reach the maximum value $V_2(\phi)=V_0\alpha$. In Fig.~(\ref{fig:1-phantom}), we plot the evolution of density parameter $\Omega_\phi$ in the scenario of phantom. Notice that the energy density in the phantom field becomes cosmologically dominant only in the recent past. In the future, the field comes to rest at the maximum of the potential and the universe eventually settles into a de Sitter phase.

\subsection{Model 2}

Write down the EoS of the scalar fields for model 2:

\begin{equation}
 \omega_{de}=\frac{ P_{scalar field}}{\rho_{scalar field}}=-1-\frac{\frac{1}{3}\delta (1+z)^{-1}}{\gamma+\delta (1+z)^{-1}} \label{eq3-30}.
 \end{equation}

It is obvious that only when parameters ($\gamma$, $\delta$) have opposite signs, there will be a singularity coming out when $z=-1-\frac{\delta}{\gamma}$. In section IV, data fitting results will show that such a singularity would not appear at low shift.

In the quintessence scenario, compare Eq.~(\ref{eq3-13}) and Eq.~(\ref{eq3-14}) with Eq.~(\ref{eq1-12}) and Eq.~(\ref{eq1-11}), we can obtain
\begin{eqnarray}
 -P_c-\frac{3}{4}P_d a &=&  \frac{1}{2}\dot{\phi}^2+V_1(\phi) \label{eq3-31} ,\\
P_c+P_d a &=&  \frac{1}{2}\dot{\phi}^2-V_1(\phi) \label{eq3-32}.
\end{eqnarray}

Simplify the above two equations, referring to Eqs.~(\ref{eq1-5})~(\ref{eq1-6})~(\ref{eq1-13})~(\ref{eq1-14})~(\ref{eq1-15})~(\ref{eq1-16}) replace model parameters ($P_c$ , $P_d$) with redefined parameters ($\gamma$ , $\delta$), then:

\begin{eqnarray}
\frac{1}{2}\dot{\phi}^2 &=& -\frac{1}{6}\rho_0\delta a \label{eq3-33} ,\\
 V_1(\phi) &=& \rho_0\gamma+\frac{7}{6}\rho_0\delta a  \label{eq3-34}.
\end{eqnarray}

\begin{figure}
\begin{center}

\includegraphics[width=2.6in,height=2in,angle=0]{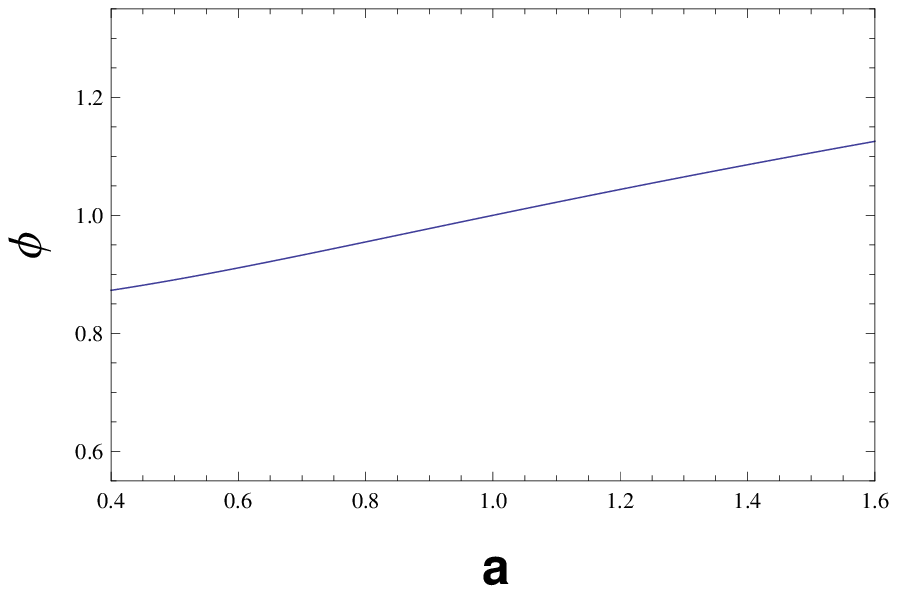}
\includegraphics[width=2.6in,height=2in,angle=0]{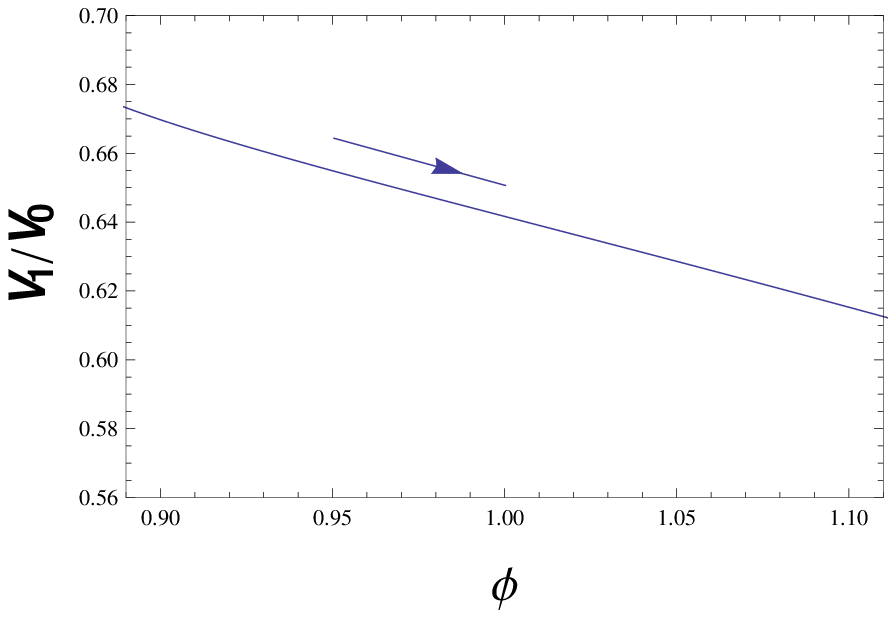}
\end{center}
\caption{The solution of Eq.~(\ref{eq3-34'}) and Eq.~(\ref{eq3-34''}) corresponding to a plus sign in Eq.~(\ref{eq3-34'}). Field $\phi$ as a function of $a$ depicted in the top panel, potential $V_1$ as a function of $\phi$ depicted in the bottom panel. The arrow indicates evolutional direction of potential with respect to time. We consider values $\gamma=0.7,\delta=-0.05$.}

\label{fig:2q1}
\end{figure}

\begin{figure}
\begin{center}
\includegraphics[width=2.6in,height=2in,angle=0]{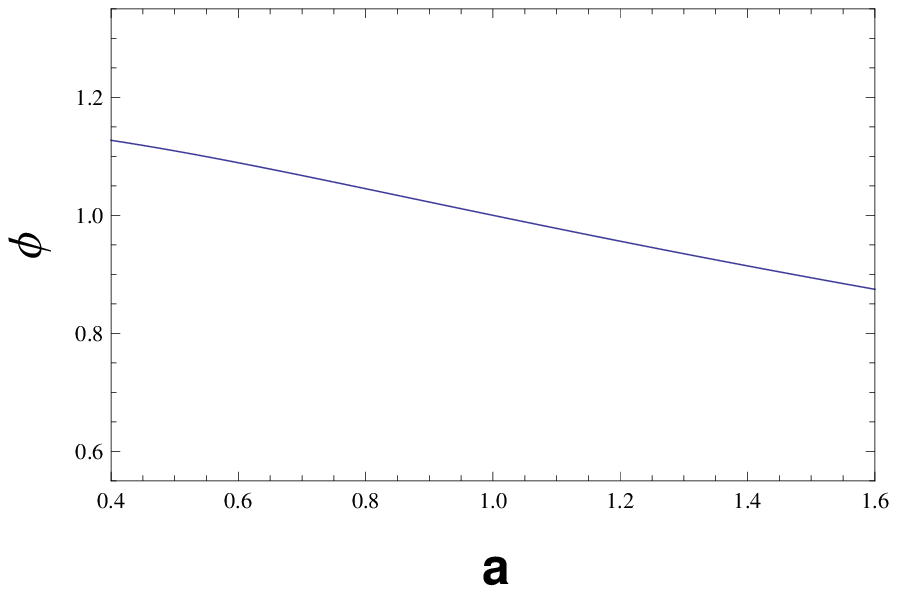}
\includegraphics[width=2.6in,height=2in,angle=0]{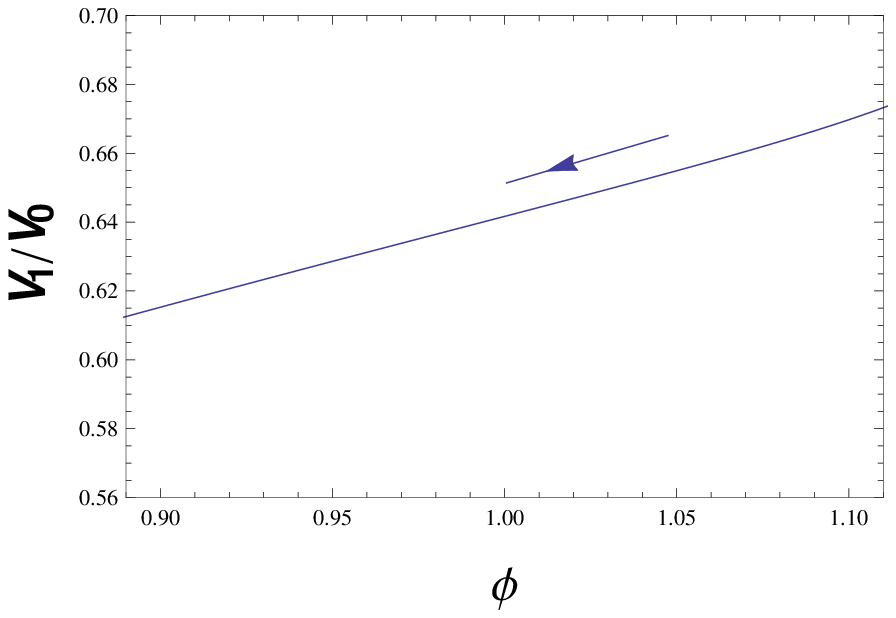}
\end{center}
\caption{The solution of Eq.~(\ref{eq3-34'}) and Eq.~(\ref{eq3-34''}) corresponding to a minus sign in Eq.~(\ref{eq3-34'}). Field $\phi$ as a function of $a$ depicted in the top panel, potential $V_1$ as a function of $\phi$ depicted in the bottom panel. The arrow indicates evolutional direction of potential with respect to time. We consider values $\gamma=0.7,\delta=-0.05$.}

\label{fig:2q2}
\end{figure}

From Eq.~(\ref{eq3-33}), it is easy to find that in the scenario of quintessence $\delta<0$. By Eq.~(\ref{eq3-33}) and Eq.~(\ref{eq3-34}), the kinetic energy $\frac{1}{2}\dot{\phi}^2$ and potential $V_1(\phi)$ of the quintessence field are constructed with parameters ($\gamma$ , $\delta$) of model 2. Simplify these two equations, we have

\begin{eqnarray}
\frac{d\phi}{da} &=& \pm M_P\sqrt{\frac{-\delta}{\gamma+\delta a+(1-\gamma-\delta) a^{-3}}} a^{-\frac{1}{2}} \label{eq3-34'},\\
 V_1(\phi) &=& V_0(\gamma+\frac{7}{6}\delta a) \label{eq3-34''},
\end{eqnarray}
where $V_0=\rho_0=3M_P^2H_0^2$. Choose parameters $\gamma=0.7,\delta=-0.05$ for numerically solving the above two equations, the two solutions are represented in Fig.~(\ref{fig:2q1}) and Fig.~(\ref{fig:2q2}) respectively. From Fig.~(\ref{fig:2q1}), we can find $\phi$ increases with $a$, and the potential decreases with the increasing $\phi$. From Fig.~(\ref{fig:2q2}), we can find $\phi$ decreases with $a$, and the potential decreases with the decreasing $\phi$. Notice that since $\delta<0$ in the scenario of quintessence, according to Eqs.~(\ref{eq1-5})~(\ref{eq1-15})~(\ref{eq3-25'}), the Friedmann equation is written as $H^2=\frac{1}{3M_P^2}\rho_0[\gamma+\delta a+(1-\gamma-\delta)a^{-3}]$, which will not hold when the scale factor $a$ is very large. Nevertheless at low redshift the relation is still feasible.

\begin{figure}[h!]
\begin{center}
\includegraphics[width=0.40\textwidth]{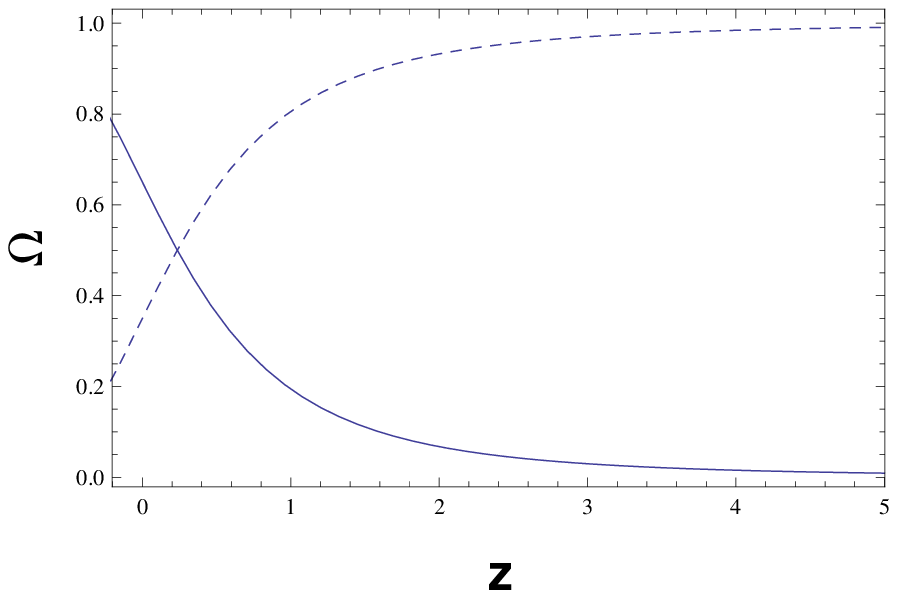}
\end{center}
\caption{Evolution of the density parameters in the quintessence field ($\Omega_\phi$) and matter ($\Omega_m$) for model 2. $\Omega_\phi$ is indicated by solid line, and $\Omega_m$ is indicated by dashed line. We consider values $\gamma=0.7,\delta=-0.05$.}
\label{fig:2-quin}
\end{figure}

By Eq.~(\ref{eq6}) and Eq.~(\ref{eq7}), we can obtain the expression of density parameter $\Omega_\phi$ for model 2:

\begin{equation}
 \Omega_\phi=\Omega_1+\Omega_2=\frac{\gamma+\delta a}{E^2} \label{eq3-34-1},
\end{equation}

\noindent
the evolution curve of density parameter $\Omega_\phi$ in the scenario of quintessence has been plotted in Fig.~(\ref{fig:2-quin}), from which we see the quintessence field begins to dominate at low redshift.

In order to realize model 2 in a phantom scenario, comparing Eq.~(\ref{eq3-18}) and Eq.~(\ref{eq3-19}) with Eq.~(\ref{eq1-12}) and Eq.~(\ref{eq1-11}), we can obtain

\begin{eqnarray}
 -P_c-\frac{3}{4}P_d a &=& -\frac{1}{2}\dot{\phi}^2+V_2(\phi) \label{eq3-35} ,\\
 P_c+P_d a &=& -\frac{1}{2}\dot{\phi}^2-V_2(\phi) \label{eq3-36}.
\end{eqnarray}

Simplify and replace model parameters ($P_c$ , $P_d$) with redefined parameters ($\gamma$ , $\delta$), we have

\begin{eqnarray}
\frac{1}{2}\dot{\phi}^2 &=& \frac{1}{6}\rho_0\delta a \label{eq3-37}, \\
 V_2(\phi) &=& \rho_0\gamma+\frac{7}{6}\rho_0\delta a \label{eq3-38}.
\end{eqnarray}

From Eq.~(\ref{eq3-37}) and Eq.~(\ref{eq3-38}), it is easy to find that in the scenario of phantom $\delta>0$. By the above two equations, one can construct the kinetic energy $\frac{1}{2}\dot{\phi}^2$ and potential $V_2(\phi)$ of the phantom field with parameters ($\gamma$ , $\delta$) of model 2. Eq.~(\ref{eq3-37}) and Eq.~(\ref{eq3-38}) can be rewritten as

\begin{equation}
\frac{d\phi}{da}=\pm M_P\sqrt{\frac{\delta}{\gamma+\delta a+(1-\gamma-\delta) a^{-3}}} a^{-\frac{1}{2}} \label{eq3-38'},
\end{equation}

\begin{equation}
 V_1(\phi)=V_0(\gamma+\frac{7}{6}\delta a) \label{eq3-38''}.
\end{equation}

\begin{figure}
\begin{center}
\includegraphics[width=2.6in,height=2in,angle=0]{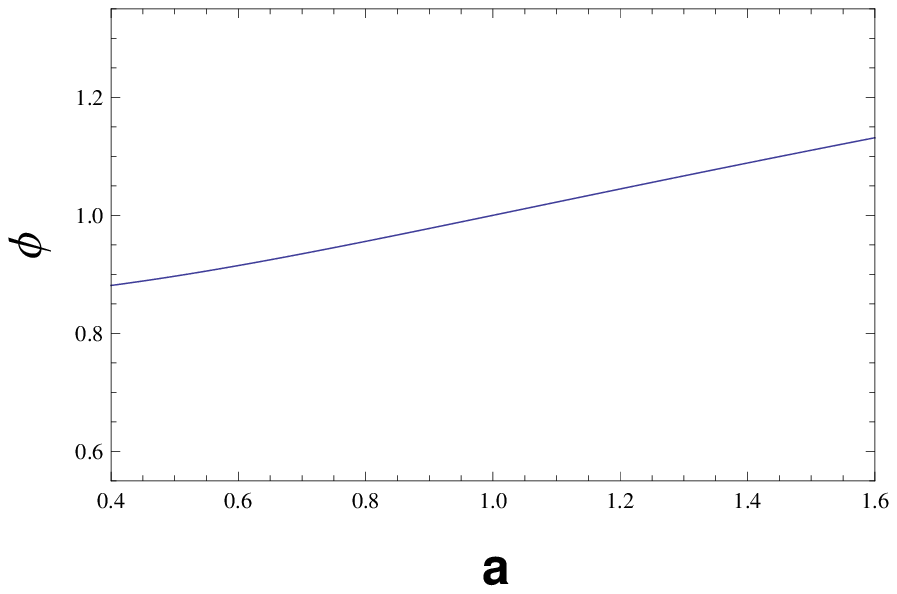}
\includegraphics[width=2.6in,height=2in,angle=0]{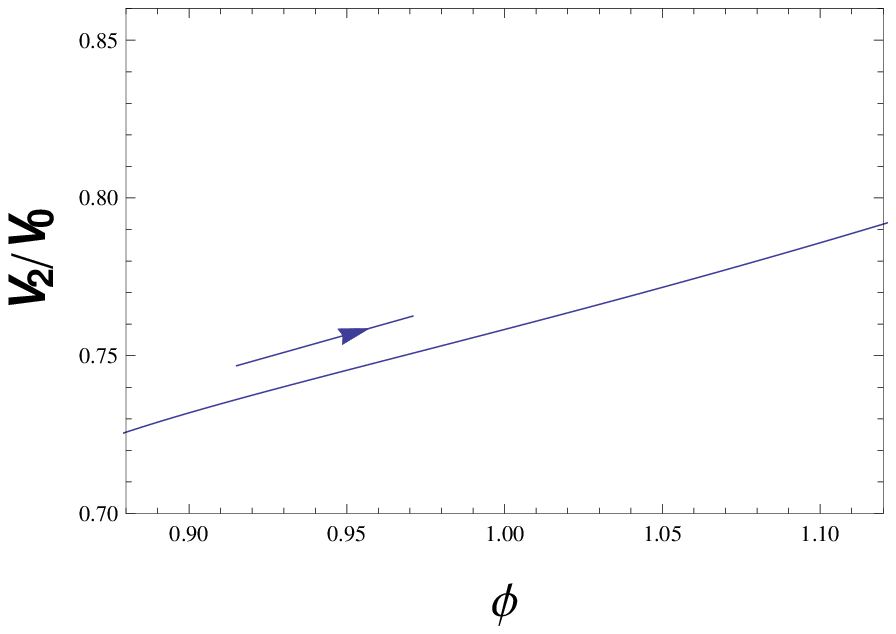}
\end{center}
\caption{The solution of Eq.~(\ref{eq3-38'}) and Eq.~(\ref{eq3-38''}) corresponding to a plus sign in Eq.~(\ref{eq3-38'}). Field $\phi$ as a function of $a$ depicted in the top panel, potential $V_2$ as a function of $\phi$ depicted in the bottom panel. The arrow indicates evolutional direction of potential with respect to time. We consider values $\gamma=0.7,\delta=0.05$.}

\label{fig:2p1}
\end{figure}

\begin{figure}
\begin{center}
\includegraphics[width=2.6in,height=2in,angle=0]{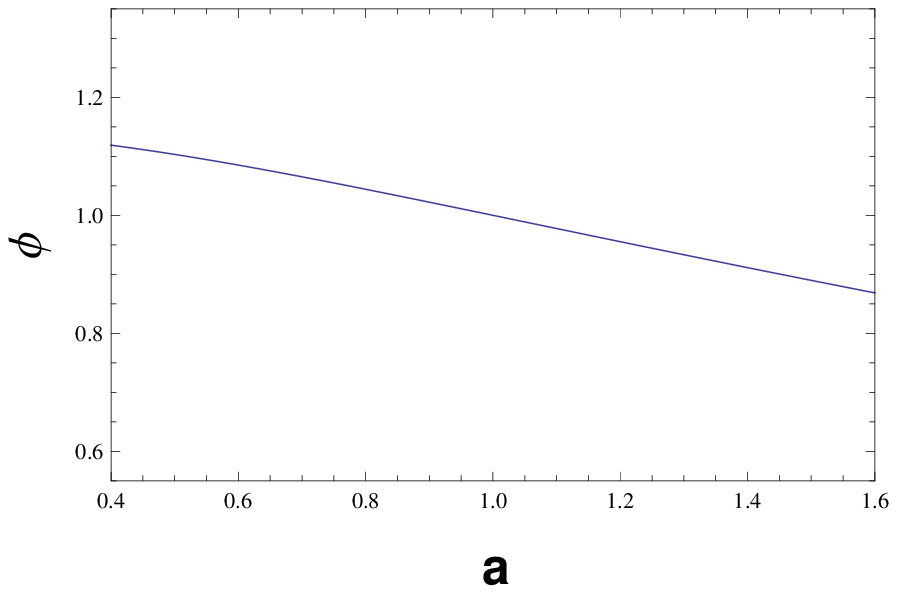}
\includegraphics[width=2.6in,height=2in,angle=0]{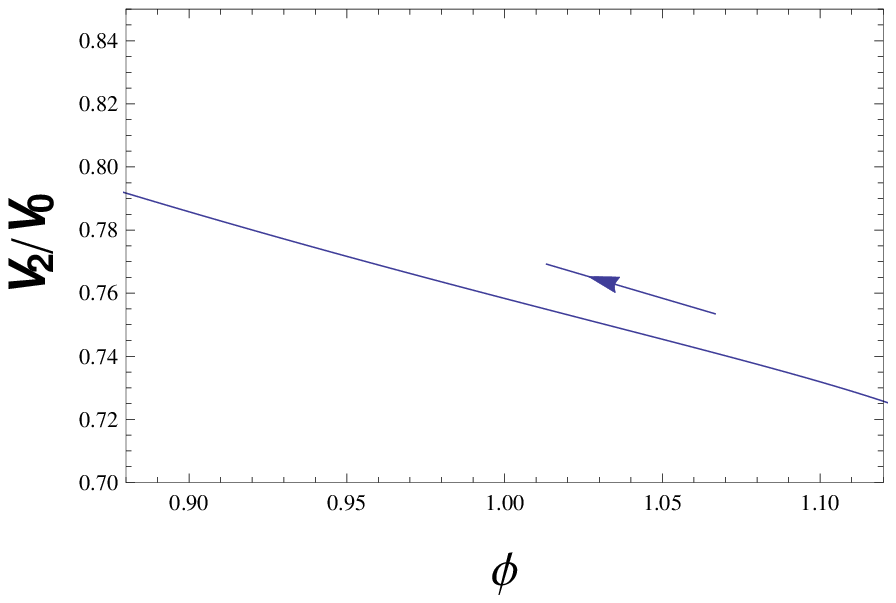}
\end{center}
\caption{The solution of Eq.~(\ref{eq3-38'}) and Eq.~(\ref{eq3-38''}) corresponding to a minus sign in Eq.~(\ref{eq3-38'}). Field $\phi$ as a function of $a$ depicted in the top panel, potential $V_2$ as a function of $\phi$ depicted in the bottom panel. The arrow indicates evolutional direction of potential with respect to time. We consider values $\gamma=0.7,\delta=0.05$. }

\label{fig:2p2}
\end{figure}

Choose parameters $\gamma=0.7,\delta=0.05$ for numerically solving the above two equations, the two solutions are represented in Fig.~(\ref{fig:2p1}) and Fig.~(\ref{fig:2p2}) respectively. From Fig.~(\ref{fig:2p1}), we can find $\phi$ increases with $a$, and the potential increases with the increasing $\phi$. In Fig.~(\ref{fig:2p2}), $\phi$ decreases with $a$, and the potential increases with the decreasing $\phi$. Notice that since $\delta>0$ in the scenario of phantom, the Friedmann equation can be written as $H^2=\frac{1}{3M_P^2}\rho_0[\gamma+\delta a+(1-\gamma-\delta)a^{-3}]$, $H\rightarrow\infty$ as $a\rightarrow\infty$, which means there will be a ``rip" in the future.

\begin{figure}[h!]
\begin{center}
\includegraphics[width=0.40\textwidth]{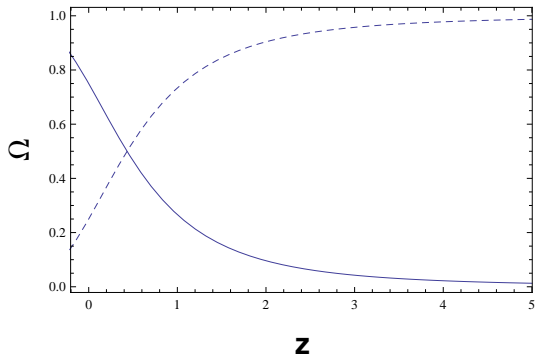}
\end{center}
\caption{Evolution of the density parameters in the phantom field ($\Omega_\phi$) and matter ($\Omega_m$) for model 2. $\Omega_\phi$ is indicated by solid line, and $\Omega_m$ is indicated by dashed line. We consider values $\gamma=0.7,\delta=0.05$.}
\label{fig:2-phantom}
\end{figure}

In Fig.~(\ref{fig:2-phantom}), we plot the evolution curve of density parameter $\Omega_\phi$ in the scenario of phantom. Note that the phantom becomes cosmologically dominant only in the recent past, finally the EoS parameter $\omega_{de}$ is less than $-1$ ($\omega_{de}=-\frac{4}{3}$, see Eq.~(\ref{eq3-30})) and the universe eventually
settles into a ``rip".

\section{Astrophysical Data Constraints}

{\bf Type Ia Supernovae}
In this paper we use the Union2.1 SNe Ia data-sets without systematic errors for data fitting, which compiles 580 SNe Ia covering the redshift range $z=[0.015,1.4]$. To perform the chi-square statistics, the theoretical
distance modulus is defined as
\begin{equation}
 \mu_{th}(z_i)\equiv 5\log_{10}D_L(z_i)+\mu_0 \label{eq2-1},
 \end{equation}
where $\mu_0\equiv 42.39-5\log_{10}h$ with $h$ the Hubble parameter in units of $100km/s/Mpc$,
\begin{equation}
    D_L= (1+z)\int^z_0\frac{dz'}{E(z';\theta)} \label{eq2-2}.
 \end{equation}
is the Hubble-free luminosity distance in a spatially flat FRW universe, $E(z;\theta)$ is the dimensionless Hubble parameter, and $\theta$ is model parameters.

 The corresponding $\chi^2_{SN}$ function is calculated from
 \begin{equation}
    \chi^2_{SN}= \sum^{580}_{i=1}\frac{[\mu_{obs}(z_i)-\mu_{th}(z_i)]^2}{\sigma^2_i} \label{eq2-3},
 \end{equation}
where $\mu_{obs}(z_i)$ and $\sigma_i$ are the observed value and the corresponding $1\sigma$ error of distance modulus for each supernova. The minimization with respect to $\mu_0$ can be made trivially by expanding $\chi^2_{SN}$ as
\begin{equation}
    \chi^2_{SN}= A-2\mu_0B+\mu^2_0C ,\label{eq2-4}
 \end{equation}
where
 \begin{eqnarray}
    A(\theta)=\sum^{580}_{i=1}\frac{[\mu_{obs}(z_i)-\mu_{th}(z_i;\theta;\mu_0=0)]^2}{\sigma^2_i}  ,\label{eq2-5} \\
    B(\theta)=\sum^{580}_{i=1}\frac{\mu_{obs}(z_i)-\mu_{th}(z_i;\theta;\mu_0=0)}{\sigma^2_i}  ,\label{eq2-6} \\
    C(\theta)=\sum^{580}_{i=1}\frac{1}{\sigma^2_i}  .\label{eq2-7}
 \end{eqnarray}

Thus $\mu_0$ is minimized as $\mu_0=\frac{B}{C}$ by calculating the following transformed $\chi^2$ :
 \begin{equation}
    \widetilde{\chi}^2_{SN}(\theta)= A(\theta)-\frac{B(\theta)^2}{C} .\label{eq2-8}
 \end{equation}

{\bf Baryon Acoustic Oscillations}
The baryon acoustic oscillation (BAO) data-sets are listed in Table~(\ref{tab1}). We use the parameter $A$ to measure the BAO peak in the distribution of SDSS luminous red galaxies. In the following $A$ is defined as
 \begin{equation}
    A\equiv \sqrt{\Omega_{m0}}E(z_b)^{-\frac{1}{3}}[\frac{1}{z_b}\int^{z_b}_0\frac{dz'}{E(z')}]^{\frac{2}{3}} ,\label{eq2-9}
 \end{equation}
where $z_b=0.35$. The $\chi^2$ for BAO data is
 \begin{equation}
    \chi^2_{BAO}= \sum^{6}_{i=1}\frac{[A_{obs}(z_i)-A_{th}(z_i;\theta)]^2}{\sigma^2_A} .\label{eq2-10}
 \end{equation}

The total $\chi^2$ is given by
\begin{equation}
    \chi^2 = \widetilde{\chi}^2_{SN}+\chi^2_{BAO} .\label{eq2-11}
 \end{equation}

\begin{figure}[h!]
\begin{center}
\includegraphics[width=0.3\textwidth]{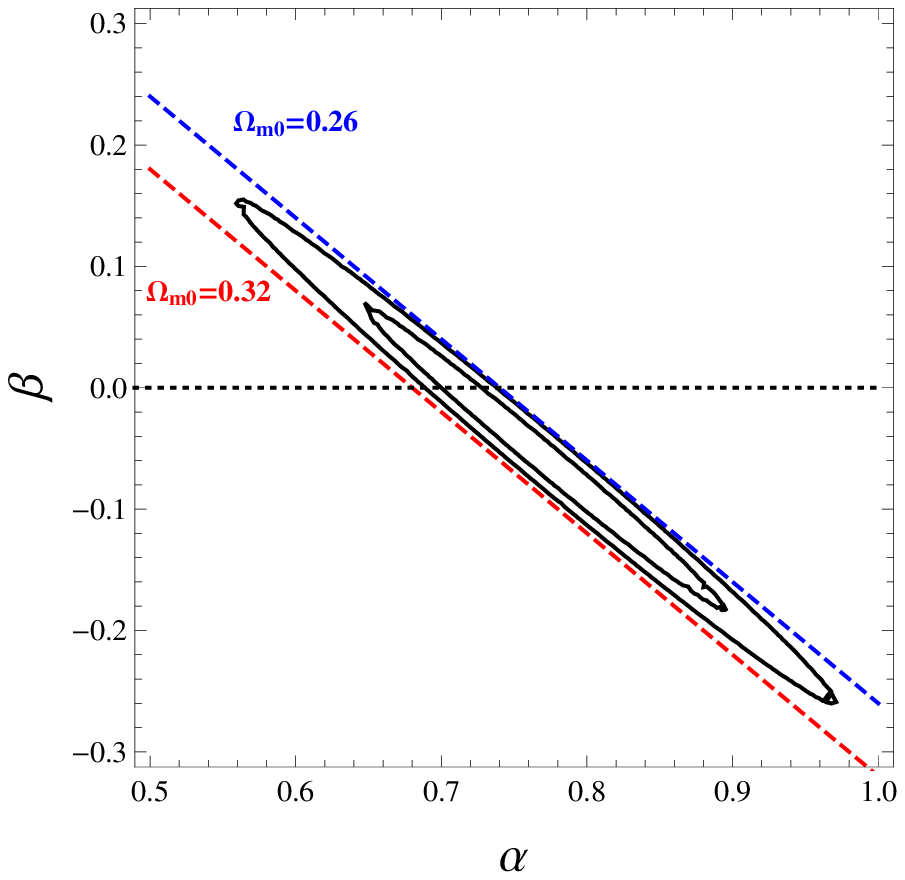}
\end{center}
\caption{$1\sigma$ and $2\sigma$ confidence ranges for parameter pair $(\alpha,\beta)$ of model 1, constrained by SNe Ia and BAO data-sets. The dotted straight line($\beta=0$) corresponds to a $\Lambda$CDM model. The blue dotted line and red dotted line correspond to $\Omega_{m0}=0.26$ and $\Omega_{m0}=0.32$, respectively.}
\label{fig:CR}
\end{figure}

\begin{figure}[h!]
\begin{center}
\includegraphics[width=0.3\textwidth]{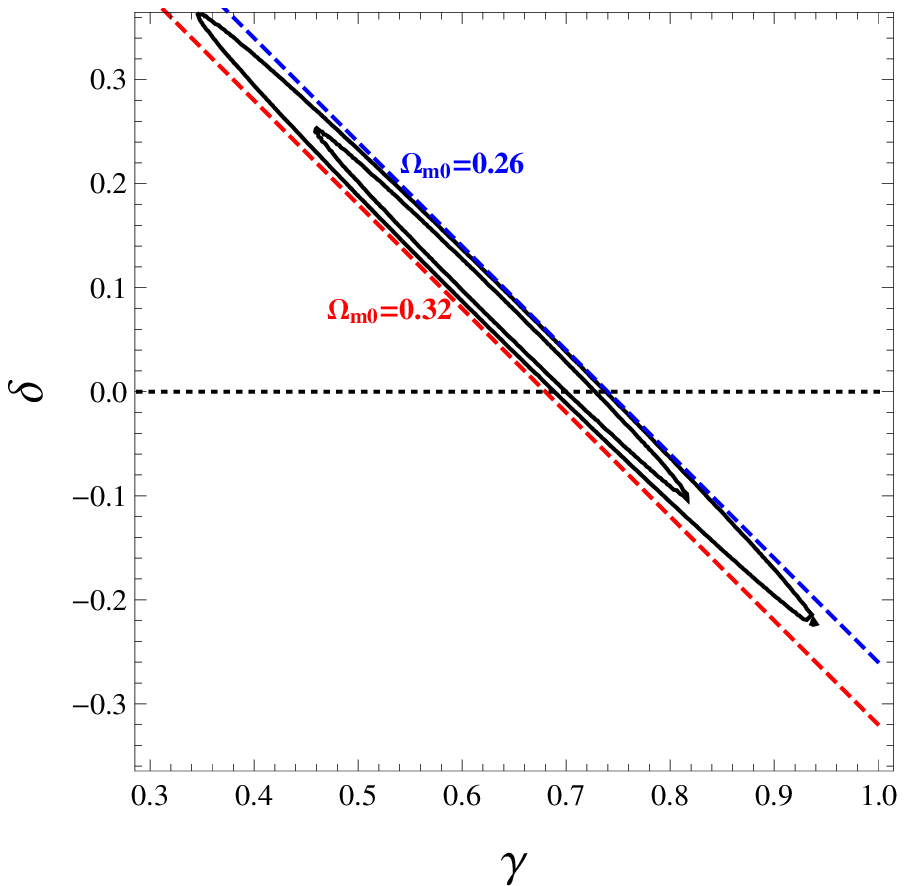}
\end{center}
\caption{$1\sigma$ and $2\sigma$ confidence ranges for parameter pair $(\gamma,\delta)$ of model 2, constrained by SNe Ia and BAO data-sets. The dotted straight line($\delta=0$) corresponds to a $\Lambda$CDM model. The blue dotted line and red dotted line correspond to $\Omega_{m0}=0.26$ and $\Omega_{m0}=0.32$, respectively.}
\label{fig:CR2}
\end{figure}

\begin{figure}[h!]
\begin{center}
\includegraphics[width=0.4\textwidth]{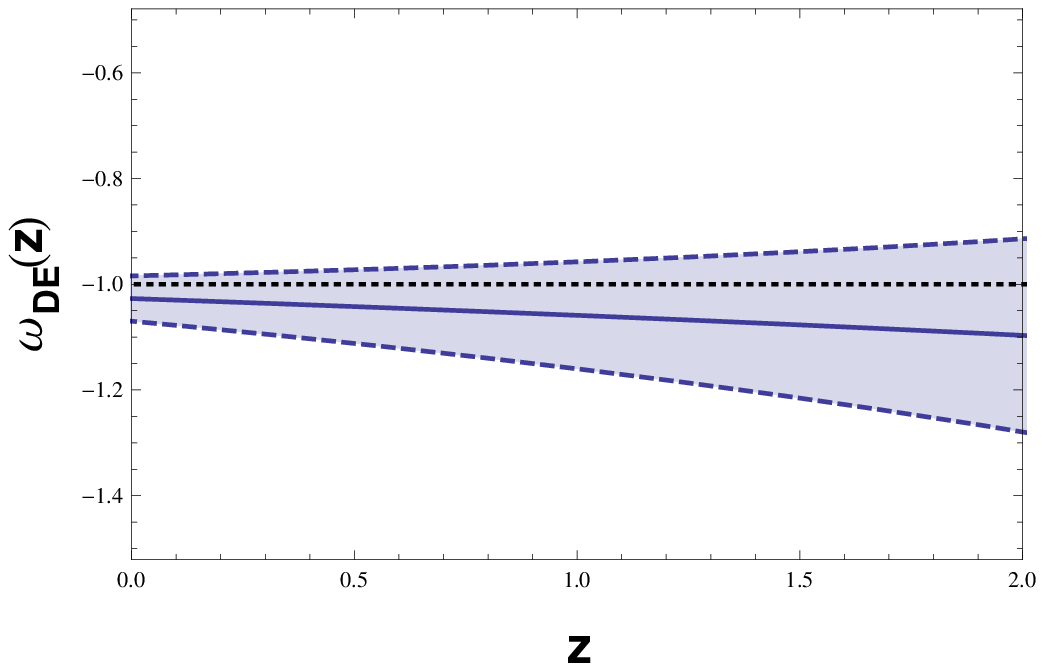}
\end{center}
\caption{Evolution of the EoS parameter $\omega_{de}$ as a function of the redshift $z$ with $1\sigma$ error propagation, constrained by SNe Ia and BAO data-sets for model 1. The solid line, the straight dotted line, and  light blue region represent the best-fit, $\omega_{de}=-1$($\Lambda$CDM), and $1\sigma$ region, respectively. }
\label{fig:w1}
\end{figure}

\begin{figure}[h!]
\begin{center}
\includegraphics[width=0.4\textwidth]{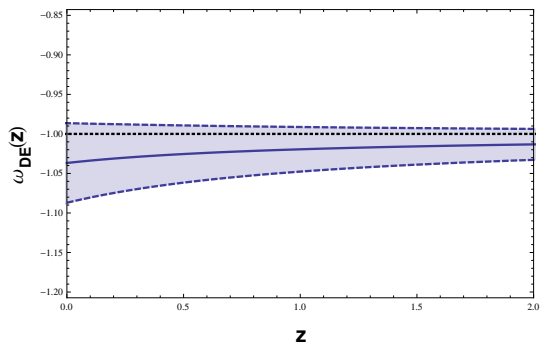}
\end{center}
\caption{Evolution of the EoS parameter $\omega_{de}$ as a function of the redshift $z$ with $1\sigma$ error propagation, constrained by SNe Ia and BAO data-sets for model 2. The solid line, the straight dotted line, light blue region represent the best-fit, $\omega_{de}=-1$($\Lambda$CDM), and $1\sigma$ region, respectively.}
\label{fig:w2}
\end{figure}

The fitting results and corresponding reduced $\chi^2$ for model 1 and model 2 are listed in Table~(\ref{tab2}). The likelihoods of parameter ($\alpha$ , $\beta$) and ($\gamma$ , $\delta$) are shown in Fig.~(\ref{fig:CR}) and Fig.~(\ref{fig:CR2}), respectively. Besides, evolution of the EoS parameter $\omega_{de}$ with respect to redshift z with $1\sigma$ error propagation from data-fitting (Tab.~\ref{tab2}) are shown in Fig.~(\ref{fig:w1}) and Fig.~(\ref{fig:w2}), respectively.

\begin{table}[h!]
\caption{6 measurement points of the Baryon Acoustic Oscillation Data-sets.}
\label{tab1}
\begin{tabular}{cccc}
\hline\noalign{\smallskip}
    redshift &  $\mathcal{A}$ & $\sigma_{\mathcal A}$ & Sample \\
\hline
0.106&	0.526&	0.028& 	6dFGS~\cite{bao1}\\
0.20&		0.488&	0.016&	SDSS~\cite{bao1}\\
0.35&		0.484&	0.016& 	SDSS~\cite{bao1}\\
0.44&		0.474&	0.034&	WiggleZ~\cite{bao1}\\
0.6&		0.452&	0.018&	WiggleZ~\cite{bao1}\\
0.73&		0.424&	0.021& 	WiggleZ~\cite{bao1}\\
\noalign{\smallskip}\hline
\end{tabular}
\end{table}

\begin{table}[h!]
\caption{parameters of model 1 and model 2 estimated by SNe Ia and BAO data-sets with $1\sigma$ errors.}
\label{tab2}
\begin{tabular}{cccc}
\hline
model 1& &model 2&\\
\hline
$\chi^2_{min}/d.o.f.$ & $564.045/(583)$&$\chi^2_{min}/d.o.f.$ &$564.098/(583)$\\
$\alpha$  & $0.771\pm 0.084$ & $\gamma$ &$0.635\pm 0.119$\\
$\beta$  & $-0.058\pm 0.084$& $\delta$ &$0.079\pm 0.120$\\
\hline
\end{tabular}
\end{table}

\section{Conclusion}

Since the observational confirmation on late-stage accelerative expansion of the universe many years ago, different models have been proposed to explain its source, among which parameterization is a widely used scheme to better characterize the dark energy with observational results. In this paper, we studied two models parameterizing the effective pressure at low redshift, $P(z)=P_a+P_b z$ and $P(z)=P_c+\frac{P_d}{1+z}$.

Deviations from the $\Lambda$CDM can be realized through different physical scenarios. Roughly speaking, there are two ways. One is to introduce some small but nonzero components besides the cosmological constant $\Lambda$, such as imperfect fluid cosmology~\cite{xh1,xh2,xh3,xh4} and cosmic strings~\cite{Sumit,Nemiroff}; whereas the other is to assume the cosmological constant $\Lambda$ exactly zero and the dark energy characterized by scalar fields evolving with time. In this paper, we pick the second way. We presented two parameterizations in the scenarios of quintessence and phantom fields, and accordingly expressed the kinetic energy term $\frac{1}{2}\dot{\phi}^2$ and potential term $V(\phi)$ with model parameters ($\alpha$ , $\beta$) and ($\gamma$ , $\delta$) respectively. Then we reconstruct the density parameter $\Omega_\phi$ for quintessence and phantom evolving with redshift. In order to obtain a better physical understanding of the field and potential, we numerically solved the field as a function of the scale factor $a$ and the potential as a function of field $\phi$.

We constrained model parameters ($\alpha$ , $\beta$) and ($\gamma$ , $\delta$) with SNe Ia and BAO data-sets. We reconstructed evolution of the EoS parameter $\omega_{de}$ in term with the redshift $z$. For model 1, the value for EoS parameter $\omega_{de0} $ is $-1.027^{0.043}_{ -0.043}$ at present day; for model 2, $\omega_{de0} =-1.037^{+0.050}_{-0.050}$. These results show that model 1 and model 2 both slightly indicate that the EoS parameter of dark energy $\omega_{de}<-1$, which corresponds to a phantom dark energy scenario at present. Still, we can not rule out a quintessence dark energy scenario or a $\Lambda$ dark energy scenario.

Different parameterizations possess their own advantages in addressing some particular problems, but their validity may not be ensured when applied for explaining global evolution. For example, our two parameterizations on effective pressure can estimate the deviation from the prediction of standard model at low redshift with generality that does not depend on the concrete physical mechanism behind.

\section*{Acknowledgement}
We are grateful for Jiaxin Wang's instruction on model building and data analyzing. We also appreciate Prof. S.~D.~Odintsov's recommendation of Refs.~\cite{modifiedgravity00,modifiedgravity000,phantom5,modifiedgravity0}and Prof. V.~K.~Onemli's recommendation of Refs.~\cite{phantom1',phantom1'',phantom1'''}.

\end{document}